\documentclass[10pt,conference]{IEEEtran}
\usepackage{graphicx,cite,amssymb,amsmath}
\usepackage{amsthm}
\usepackage{caption}
\usepackage{subcaption}
\usepackage{bbm}
\usepackage{algorithm}
\usepackage{algorithmic}
\usepackage{multirow}
\usepackage{geometry}

\newtheorem{theorem}{Theorem}

\newtheorem{remark}{Remark}

\IEEEoverridecommandlockouts

\begin{document}


\newgeometry{top=0.55in, bottom=0.55in, left=0.75in, right=0.75in}

\title{\huge{Compressive Channel Estimation and Multi-user Detection \\ in C-RAN }
\author{
\IEEEauthorblockN{Qi~He$^{\dag\S}$, Tony~Q.S.~Quek$^{\S}$, Zhi~Chen$^{\dag}$ and Shaoqian Li$^{\dag}$
\\
\IEEEauthorblockA{$^{\dag}$\,National Key Laboratory on Communications, University of Electronic Science and Technology of China\\
$^\S$\,Information Systems Technology and Design Pillar, Singapore University of Technology and Design\\
heqi.tech@hotmail.com; tonyquek@sutd.edu.sg; chenzhi@uestc.edu.cn; lsq@uestc.edu.cn
}
}}
}

\maketitle

\begin{abstract}
This paper considers the channel estimation (CE) and multi-user detection (MUD) problems in cloud radio access network (C-RAN).
Assuming that active users are sparse in the network, we solve CE and MUD problems with compressed sensing (CS) technology to greatly reduce the long identification pilot overhead.
A mixed $\ell_{2,1}$-regularization functional for extended sparse group-sparsity recovery is proposed to exploit the inherently sparse property existing both in user activities and remote radio heads (RRHs) that active users are attached to.
Empirical and theoretical guidelines are provided to help choosing tuning parameters which have critical effect on the performance of the penalty functional.
To speed up the processing procedure, based on alternating direction method of multipliers and variable splitting strategy, an efficient algorithm is formulated which is guaranteed to be convergent.
Numerical results are provided to illustrate the effectiveness of the proposed functional and efficient algorithm.
\end{abstract}


\section{Introduction}


The cloud radio access network (C-RAN) has been proposed as a breakthrough network architecture to improve spectral efficiency and energy efficiency \cite{mobile2011c,ZTE2011}.
In C-RAN, strong functional but high-power-consumption base stations (BSs) are decoupled into two parts: distributed remote radio heads (RRHs) and baseband units (BBUs) \cite{checko2015cloud}.
The BBUs are consolidated together to be a BBU pool
for handling all the baseband signal processing in the network.
Plenty of RRHs are distributed on a large area, and connected to BBU pool through
fronthaul links.




The estimation of channel state information (CSI) is fundamental and vital in wireless communication.
The identification of active users is also critical for resource allocation in the network.
Channel estimation (CE) and multi-user detection (MUD) problems have been investigated extensively \cite{biguesh2006training}.
However, in classical CE and MUD methods, in order to keep orthogonal characteristic between identification pilots belong to different users,
the length of identification pilots needs to be scale with the number of users times the number of antennas per user, which is a significant overhead when the network is large.
With the observation that active users are sparse,
many recent works \cite{zhu2011exploiting,shim2012multiuser,schepker2012compressive,schepker2011sparse}
on MUD and CE have taken the sparsity of user activities into consideration by
using compressed sensing (CS) technology,
which greatly reduces the identification pilot overhead.


The C-RAN architecture brings new challenges to CE and MUD.
Since the baseband processing is clustered together in C-RAN, CE and MUD is no longer restricted to local BS processing, but jointly processed
in BBU pool to meet the demand of the whole network.
Each user is linked to each RRH to enable full cooperation among RRHs.
This paper aims to detect the active users and to estimate full CSI, i.e. the channel states between each active user and each RRH in C-RAN.

%

%

Related works on CE and MUD in C-RAN can be found in \cite{xu2015active} and \cite{utkovski2016random}.
Bayesian sparsity inference tool and hybrid generalized approximate message passing method are leveraged in \cite{xu2015active} and \cite{utkovski2016random} respectively to exploit sparsity of user activities.
\cite{xu2015active} assumes multiple antennas in each RRH and user, while
\cite{utkovski2016random} adopts single-antenna RRHs and users but
takes limited capacity of fronthaul links into consideration.
Both \cite{xu2015active} and \cite{utkovski2016random} assume the Rayleigh fading channel and the exact knowledge of large-scale path loss.
However, this simplified model tends to describe static network, while mobility and environment changes introduce more uncertainty \cite{meriaux2013stochastic, de2012concurrent}.
Besides, to predict the precise large-scale path loss between each RRH and each user in C-RAN is very expensive when the network is large \cite{har1999path, jeong2001comparison}.
%
%
%
%

In this paper, we formulate a penalty functional to solve CE and MUD problems in C-RAN.
Our formulation is general which does not depend on the prior information of channel parameters or the assumption of channel model.
We only assume the sparsity of active users, which is practical in real communication systems.
On the other hand, we observe that there are limited RRHs surrounding each active user in C-RAN, and the effects of most distant RRHs on this user are negligible. Hence there exist two types of sparsity: one is the sparse active users; the other is sparse RRHs related to one particular active user.
Motivated by this observation, we propose a mixed $\ell_{2,0}$-regularization functional to exploit these two types of sparsity simultaneously, and then relax it as a convex and smoothed mixed $\ell_{2,1}$-regularization functional.
A re-weighting strategy is adopted to enhance estimation accuracy.
As tuning parameters in the functional have critical influence on the performance, we also provide empirical and theoretical guidelines
for choosing the appropriate values.

Considering that the processing time of MUD and CE should be less than the channel coherence time, we propose an efficient algorithm based on alternating direction method of multipliers (ADMM) and variable splitting strategy
to reduce the computational complexity of solving the proposed functional. 
Simulation results show that, even without any prior information on the channel,
our algorithm provides almost the same performance and has almost the same computational complexity as state-of-the-art algorithm.


%
%
%
%
%
%
%

\emph{Notations}:
Uppercase (lowercase) boldface letters denote matrices (column vectors).
iff denotes ``if and only if''.
The operators $(\cdot)^T$, $(\cdot)^H$, $\left \| \cdot \right \|_F$ stand for transpose, conjugate transpose and Frobenius norm, respectively.
$\mathbf{I}_N$ denotes an $N\times N$ identity matrix.
$\otimes$ and $\circ$ denote the Kronecker product and Hadamard product (element product) respectively.
$\Re(a)$ stands for real part of complex number $a$.
$\text{vec}(\mathbf{A})$ denotes the vectorization of
$\mathbf{A}$ formed by stacking its columns into a single
column vector.
$\mathbf{A}_{[i,j]}$ stands for the $\{i,j\}$th element in $\mathbf{A}$, and
$\mathbf{A}_{i,j}$ for the $\{i,j\}$th submatrix in $\mathbf{A}$.
$\mathbf{A}^{\circ(a)}$ denotes an element-wise exponentiation operator on $\mathbf{A}$, i.e., $(\mathbf{A}^{\circ(a)})_{[i,j]}=(\mathbf{A}_{[i,j]})^{a}$.


\section{System Model}
We consider a C-RAN architecture with $G$ RRHs and $K$ users. There are $M$ antennas in each RRH, and $N$ antennas in each user device.
The length of identification pilots is $L$, and the $k$th user is assigned with a pilot matrix $\mathbf{P}_{k} \in \mathbb{C}^{N\times L}$.
We assume that only a small part of users are active.
The set of active users is indicated by $\mathcal{A} \subsetneqq  \{1,...,K\}$.
An indicator function is defined to mark whether the $k$th user is active,
\begin{equation}
\mathbbm{1}_\mathcal{A}(k)=\left\{\begin{matrix}1,\  \text{if}\ k \in \mathcal{A}
\\ 0,\  \text{if}\ k \notin \mathcal{A}
\end{matrix}\right.
\end{equation}

In the following descriptions, we use $\mathbbm{1}_k$ to denote $\mathbbm{1}_\mathcal{A}(k)$ if there is no confusion.

The quasi-static channel between $k$th user and $g$th RRH is denoted by complex matrix $\mathbf{H}_{g,k} \in \mathbb{C}^{M\times N}$.
As only active users transmit their identification pilots, the received pilot data in the $g$th RRH can be described as
\begin{equation}
\mathbf{R}_g=\sum_{k\in \mathcal{A}}\mathbf{H}_{g,k}\mathbf{P}_{k} +\bar{\mathbf{N}}_g
=\sum_{k=1}^K\mathbf{H}_{g,k}\mathbbm{1}_k\mathbf{P}_{k}+\bar{\mathbf{N}}_g,\ g = 1,... ,G\
\end{equation}
where $\bar{\mathbf{N}}_g \in \mathbb{C}^{M\times L}$ denotes the antenna additive noise in the $g$th RRH.

By stacking received pilot data in all the $G$ RRHs, the data received in BBU can be described as
\begin{equation}
\label{b4}
\mathbf{R}=\mathbf{H}\mathbf{\Lambda}\mathbf{P}+\bar{\mathbf{N}}
\end{equation}
where $\mathbf{R}=[\mathbf{R}^T_1,\ ...\ ,\mathbf{R}^T_G]^T$,
$\mathbf{\Lambda}=\text{diag}[\mathbbm{1}_1,\ ...\ ,\mathbbm{1}_K]\otimes \mathbf{I}_N$,
$\mathbf{P}=[\mathbf{P}^T_1,\ ...\ ,\mathbf{P}^T_K]^T$,
$\bar{\mathbf{N}}=[\bar{\mathbf{N}}^T_1,\ ...\ ,\bar{\mathbf{N}}^T_G]^T$,
and
\begin{equation}
\mathbf{H} =
\begin{bmatrix}
\mathbf{H}_{1,1} & \cdots & \mathbf{H}_{1,K} \\
\cdots & \cdots & \cdots
\\ \mathbf{H}_{G,1} & \cdots & \mathbf{H}_{G,K}
\end{bmatrix}
\end{equation}

Besides the received pilot data $\mathbf{R}$, BBU also has the knowledge of identification pilot matrix $\mathbf{P}$.
Hence, our MUD problem is to estimate $\mathbf{\Lambda}$ in (\ref{b4}).
With the constraint of sparse user activities, our CE problem is to estimate matrices $[\mathbf{H}^T_{1,k};...;\mathbf{H}^T_{G,k}]^T$, $\forall k \in \mathcal{A}$.

\section{Problem Formulations}
By taking the transposition of equation (\ref{b4}), we get a compressed sensing problem as below.
\begin{equation}
\label{c1}
\mathbf{R}^H=\mathbf{P}^H\mathbf{\Lambda}\mathbf{H}^H+\bar{\mathbf{N}}^H
\end{equation}
where $\mathbf{P}^H$ is a fat matrix with the assumption that the pilot length $L$ is less than the number of users $K$.
Our MUD and CE problems are converted to estimating $\mathbf{\Lambda}\mathbf{H}^H$ as a whole.

\begin{equation}
\label{c2-0}
\mathbf{\Lambda H}^H =
\begin{bmatrix}
\mathbbm{1}_1\mathbf{H}^H_{1,1} & \cdots & \mathbbm{1}_1\mathbf{H}^H_{G,1} \\
\cdots & \cdots & \cdots
\\ \mathbbm{1}_K\mathbf{H}^H_{1,K} & \cdots & \mathbbm{1}_K\mathbf{H}^H_{G,K}
\end{bmatrix}
\end{equation}



We define the submatrix
[$\mathbbm{1}_k\mathbf{H}^H_{1,k},\ ... \ ,\mathbbm{1}_k\mathbf{H}^H_{G,k}$]
as the $k$th ``row chunk'' matrix in
$\mathbf{\Lambda}\mathbf{H}^H$, and the submatrix $\mathbbm{1}_k\mathbf{H}^H_{g,k}$ as the \{\emph{k},\emph{g}\}th ``element chunk'' matrix. The sparsity of the $k$th row chunk
is determined by $\mathbbm{1}_k$, which denotes whether the $k$th user is active.
Therefore the matrix $\mathbf{\Lambda}\mathbf{H}^H$ has row-chunk sparsity structure at first.

In practical C-RAN architecture, huge numbers of RRHs are distributed on vast areas, and most of RRHs have negligible effect on a particular active user.
It gives rise to that, if the $k$th user is active, most of the element chunks in the $k$th row chunk approximate to zero, and the $k$th row chunk has sparsity structure.
Therefore, we say $\mathbf{\Lambda}\mathbf{H}^H$ has element-chunk sparsity structure.
For one particular active user, the element-chunk sparsity situation is affected by many factors, like
the position of this user, the distribution geometry of surrounding RRHs, power allocations and beamformings in RRHs, etc.

To simplify the expression, we rewrite the system model (\ref{c1}) as below,
\begin{equation}
\label{c2}
\mathbf{B}=\mathbf{A}\mathbf{X}+\mathbf{N}
\end{equation}
where $\mathbf{B}=\mathbf{R}^H \in \mathbb{C}^{L \times GM}$, $\mathbf{A}=\mathbf{P}^H\in \mathbb{C}^{L \times KN}$, $\mathbf{X}=\mathbf{\Lambda}\mathbf{H}^H \in \mathbb{C}^{KN \times GM}$, and $\mathbf{N}=\bar{\mathbf{N}}^H$.

The matrix $\mathbf{X}$ to be estimated has both row-chunk and element-chunk sparsity structures. In the following, we use $\mathbf{X}_i \in \mathbb{C}^{N \times GM}$
to denote the $i$th row chunk [$\mathbbm{1}_k\mathbf{H}^H_{1,k},\ ... \ ,\mathbbm{1}_k\mathbf{H}^H_{G,k}$],
and $\mathbf{X}_{i,j}\in \mathbb{C}^{N \times M}$
to denote the $\{i,j\}$th element chunk $\mathbbm{1}_i\mathbf{H}^H_{i,j}$.
We define $\mathbf{A}_{i} = \mathbf{P}_i^T \in \mathbb{C}^{L \times N}$ in (\ref{c1}), then $\mathbf{A} = [\mathbf{A}_{1},\ ... \ ,\mathbf{A}_{K}]$;
and define $\mathbf{B}_{i} = \mathbf{R}_i^T \in \mathbb{C}^{L \times M}$ in (\ref{c1}), then $\mathbf{B} = [\mathbf{B}_{1},\ ... \ ,\mathbf{B}_{G}]$.

To exploit row-chunk and element-chunk sparsity structures simultaneously,
we propose a mixed $\ell_{2,0}$-regularization functional shown as blow,
\begin{equation}
\label{c3-0}
\min\limits_{\mathbf{X}}\ \alpha_1 \sum_{i=1}^K \mathbbm{1}(\mathbf{X}_i )
+ \alpha_2\sum_{i=1}^K \sum_{j=1}^G \mathbbm{1}(\mathbf{X}_{i,j})
+\frac{1}{2}\left \| \mathbf{A}\mathbf{X}-\mathbf{B} \right \|^2_F
\end{equation}
where $\mathbbm{1}(\cdot)$ is an indicator function defined as following,
\begin{equation}
\mathbbm{1}(\mathbf{D})=\left\{\begin{matrix}1,\ \text{if}\ \mathbf{D}\text{ is a non-zero matrix}
\\  0,\ \text{if}\ \mathbf{D}\text{ is a zero matrix} \hfill
\end{matrix}\right.
\end{equation}

In the functional (\ref{c3-0}), the first term exploits the row chunk sparsity, and the second term takes into account the element chunk sparsity.
The absolute values of tuning parameters $\alpha_1$ and $\alpha_2$ control
the tradeoff between the sparsity of the solution and the quality of fit, while
the relative values between $\alpha_1$ and $\alpha_2$ control the balance of row-chunk and element-chunk sparsity.

As (\ref{c3-0}) is non-smooth, we will relax it to a smooth and convex mixed $\ell_{2,1}$-regularization functional.
However, there is one drawback in $\ell_{2,1}$-regularization functionals that coefficients with larger magnitudes are penalized more heavily than coefficients with smaller magnitudes,
which limits the performance and also exhibits a key difference between $\ell_{2,1}$ and $\ell_{2,0}$ regularizations.

An iteratively re-weighted $\ell_{1}$-norm minimization algorithm was proposed
in \cite{candes2008enhancing} to provide approximative ``democratic penalization'' on coefficients and further enhance the performance of $\ell_{1}$-norm functional.
Similarly to that, we introduce the weight matrix $\mathbf{W}$, and propose a weighted $\ell_{2,1}$-regularization functional as below,
\begin{align}
\label{c3}
\min\limits_{\mathbf{X}}\ \alpha_1 \sum_{i=1}^K \left \| \mathbf{W}_i \circ \mathbf{X}_i \right \|_F
&+ \alpha_2 \sum_{i=1}^K \sum_{j=1}^G \left \| \mathbf{W}_{i,j}\circ \mathbf{X}_{i,j} \right \|_F \nonumber \\
&+\frac{1}{2} \left \| \mathbf{A}\mathbf{X}-\mathbf{B} \right \|^2_F
\end{align}
where
$\mathbf{W}_{i}$ and $\mathbf{W}_{i,j}$ denote the $i$th row-chunk submatrix and the $\{i,j\}$th element-chunk submatrix in $\mathbf{W}$ respectively.

The functional (\ref{c3}) is in the context of multiple measurement vectors (MMV) scenario.
If we suppose that each RRH and user owns only one antenna, i.e., $M=N=1$, then each element chunk $\mathbf{X}_{i,j}$ reduces to a complex number.
By letting $\mathbf{b}=\text{vec}(\mathbf{B}^T)$, $\mathbf{x}=\text{vec}(\mathbf{X}^T)$, $\mathbf{n}=\text{vec}(\mathbf{N}^T)$ and $\hat{\mathbf{A}}=\mathbf{A}\otimes \mathbf{I}_G$,
from equation (\ref{c2}) we get
\begin{equation}
\label{c3-1}
\mathbf{b}=\hat{\mathbf{A}}\mathbf{x}+\mathbf{n}
\end{equation}
which is a single measurement vector (SMV) model.
The vector $\mathbf{x}$ to be estimated has two types of sparsity: ``group-wise sparsity'' and ``element-wise sparsity within group'', where the group size is $G$.
By removing the effect of weight matrix $\mathbf{W}$,
our proposed functional (\ref{c3}) is reduced to a sparse group lasso criterion \cite{simon2013sparse} as following,
\begin{align}
\label{c3-2}
\min\limits_{\mathbf{x}}\ \alpha_1 \sum_{i=1}^K \left \| \mathbf{x}^{(i)} \right \|_F
+ \alpha_2 \left \| \mathbf{x} \right \|_1 +\frac{1}{2} \left \| \hat{\mathbf{A}}\mathbf{x}-\mathbf{b} \right \|^2_F
\end{align}
where $\mathbf{x}^{(i)}$ denotes the $i$th group in $\mathbf{x}$, and $\left \| \mathbf{x} \right \|_1$ denotes the sum of magnitudes of each element in $\mathbf{x}$.

Considering feasibility of multiple antennas in RRHs and users,
and also higher estimation accuracy demand, we adopt the functional (\ref{c3}), which can be viewed as extended version of sparse group-sparsity recovery problem.
If only single-antenna RRHs and user devices are deployed in the C-RAN architecture, (\ref{c3-2}) can be used to get normal precise estimation, and can be solved by several existing algorithms \cite{simon2013sparse}, \cite{foygel2010exact}, etc.

Next, we discuss the choices of $\alpha_1$ and $\alpha_2$,
which have critical effect on the performance of (\ref{c3}).
Smaller values of $\alpha_1$ and $\alpha_2$ may be insufficient to recover the
interested sparse signal, and larger values may lead to biased estimation.
However, to the best of our knowledge, determining
proper values for tuning parameters like $\alpha_1$ and $\alpha_2$ still remains an implementation-level issue.

\begin{remark}
\label{remark-1}
The choice of $\alpha_1$ and $\alpha_2$ are empirical.
In general, a larger $\alpha_1$ should be chosen when the C-RAN is small and dense, where
the row-chunk sparsity dominates and the element-chunk sparsity is non-significant.
When the C-RAN is large, we empirically choose a larger $\alpha_2$ to exploit the element-chunk sparsity efficiently.
\end{remark}

Here we further provide rigorously numerical upper bounds to help choosing $\alpha_1$ and $\alpha_2$.
We notice that if $\alpha_1=0$ or $\alpha_2=0$, only one type sparsity is grasped, and (\ref{c3}) reduces to an extension of group-lasso functional \cite{meier2008group}.
Inspired by this insight, we further arrive at the following result.

\begin{theorem}
\label{Theorem1}
The solution of the functional (\ref{c3}) is $\hat{\mathbf{X}}=\mathbf{0}$ if
$\alpha_1\geq \alpha_1^*=\max\limits_{i} \left \|   (\mathbf{A}^H_{i}\mathbf{B})\circ (\mathbf{W}_i)^{\circ(-1)}\right \|_F$
or
$\alpha_2\geq \alpha_2^*=\max\limits_{i,j} \left \|  (\mathbf{A}^H_{i}\mathbf{B}_{j}) \circ (\mathbf{W}_{i,j})^{\circ(-1)}\right \|_F$.

\end{theorem}


The proof of Theorem \ref{Theorem1} can be found in Appendix A.
It can be seen that $\alpha_1$ should be chosen strictly less than $\alpha_1^*$, and $\alpha_2$ less than $\alpha_2^*$,
to prevent identically zero solution.



\section{Algorithm Design}

\subsection{Standard Second-Order Cone Programming}
The proposed functional (\ref{c3}) can be transformed into a second-order cone programming (SOCP) problem as below, and solved by standard method, such as interior point method \cite{boyd2004convex}.
\begin{align}
\label{d0}
&\min\limits_{\mathbf{X}}\ \alpha_1 \sum_{i=1}^K \mathbf{c}_i
+ \alpha_2\sum_{i=1}^K \sum_{j=1}^G \mathbf{D}_{i,j}
+ e \nonumber \\
&\text{ s.t. }
\left \| \mathbf{W}_i \circ \mathbf{X}_i \right \|_F \leq \mathbf{c}_i \nonumber \\
&\ \ \ \ \ \left \| \mathbf{W}_{i,j}\circ \mathbf{X}_{i,j} \right \|_F \leq \mathbf{D}_{i,j} \nonumber \\
&\ \ \ \ \ \ \frac{1}{2} \left \| \mathbf{A}\mathbf{X}-\mathbf{B} \right \|^2_F \leq e
\end{align}
where $\mathbf{c}\in \mathbb{R}_+^{K\times 1}$ and $\mathbf{D}\in \mathbb{R}_+^{K\times G}$.


The SOCP method is guaranteed to find the optimal solution of our convex functional (\ref{c3}) with fixed $\mathbf{W}$.
We iterate the procedure (\ref{d0}) for several times, in each of which we use previous estimation of $\mathbf{X}$ to determine $\mathbf{W}$ similarly to \cite{candes2008enhancing}. As an empirical law, each element in $\mathbf{W}$ is set inversely to previous estimation of element magnitude, as below
\begin{align}
\label{d11}
\mathbf{W}_{[i,j]}=\frac{1}{abs\big(\mathbf{X}_{[i,j]}\big)+\epsilon}
\end{align}
where $\epsilon > 0$ is a very small positive number to provide stability.

\subsection{Alternating Direction Method of Multipliers}
Solving SOCP problem in (\ref{d0}) has high computational complexity when the problem size is medium to large scale.
In practical C-RAN system, the huge number of users and RRHs make it impractical to solve (\ref{c3}) by standard algorithm like SOCP.
Given that, we propose an efficient algorithm to accelerate the problem-solving procedure.
The alternating direction method of multipliers (ADMM) combined with variable splitting strategy can be used to separate our problem into simple sub-problems. Methods based on ADMM are easy to implement and also own a guaranteed convergency property \cite{boyd2011distributed}.



We first split the variable $\mathbf{X}$ in (\ref{c3}), and bring two auxiliary variables $\mathbf{Z}$ and $\mathbf{Q}$:
\begin{align}
\label{d1}
\min\limits_{\mathbf{X},\mathbf{Z},\mathbf{Q}} \alpha_1 \sum_{i=1}^K \left \| \mathbf{Z}_i \right \|&_F
+ \alpha_2\sum_{i=1}^K \sum_{j=1}^G \left \| \mathbf{Q}_{i,j} \right \|_F
+\frac{1}{2} \left \| \mathbf{A}\mathbf{X}-\mathbf{B} \right \|^2_F\
\nonumber   \\
&\text{s.t.} \  \mathbf{Z}=\mathbf{W}\circ \mathbf{X},\ \mathbf{Q}=\mathbf{W}\circ \mathbf{X}
\end{align}

As (\ref{d1}) is separable in variables $\mathbf{Z}$, $\mathbf{Q}$ and $\mathbf{X}$, ADMM is applicable.
We transform (\ref{d1}) to be an augmented Lagrangian problem as below,
\begin{align}
\centering
\label{d2}
\mathcal{L}&(\mathbf{X},\mathbf{Z},\mathbf{Q},\boldsymbol{\lambda}_1,\boldsymbol{\lambda}_2)\nonumber   \\
=&\ \alpha_1 \sum_{i=1}^K \left \| \mathbf{Z}_i \right \|_F
+ \alpha_2\sum_{i=1}^K \sum_{j=1}^G \left \| \mathbf{Q}_{i,j} \right \|_F
+\frac{1}{2} \left \| \mathbf{A}\mathbf{X}-\mathbf{B} \right \|^2_F\ \nonumber   \\ &-\Re\Big{\{}tr\Big(\boldsymbol{\lambda}_1^H(\mathbf{Z}-\mathbf{W}\circ \mathbf{X})\Big)
+tr\Big(\boldsymbol{\lambda}_2^H(\mathbf{Q}-\mathbf{W}\circ \mathbf{X})\Big)
\Big{\}} \nonumber  \\
&+\frac{\beta}{2} \big(\left \| \mathbf{Z}-\mathbf{W}\circ \mathbf{X} \right \|^2_F
+\left \| \mathbf{Q}-\mathbf{W}\circ \mathbf{X} \right \|^2_F \big)
\end{align}
where $\boldsymbol{\lambda}_1$ and $\boldsymbol{\lambda}_2$ are Lagrange multipliers which have the same dimension as $\mathbf{X}$. $\beta$ is the  regularization parameter.

The ADMM iterates over the minimization of augmented Lagrangian problem (\ref{d2}) on variables $\mathbf{X}$, $\mathbf{Z}$ and $\mathbf{Q}$ sequentially, each of which forms a subproblem as below.


The $\mathbf{X}$-subproblem is to minimize (\ref{d2}) with respect to $\mathbf{X}$, which is a convex quadratic problem. We have the following equation by taking derivative with respect to $\mathbf{X}$ and setting it to be zero.
\begin{align}
\label{d4}
\beta\mathbf{W}\circ (\mathbf{Z}+\mathbf{Q})
+\mathbf{A}^H\mathbf{B}
&-\mathbf{W}\circ(\boldsymbol{\lambda}_1+\boldsymbol{\lambda}_2)
\nonumber   \\
=\
& 2\beta \mathbf{W}^{\circ (2)}\circ \mathbf{X}
+\mathbf{A}^H\mathbf{A}\mathbf{X}
\end{align}

By setting the left side in (\ref{d4}) equal to a newly defined matrix $\mathbf{D}$, we get the estimation of the $l$th column in $\mathbf{X}$,
\begin{align}
\label{d5}
\hat{\mathbf{X}}_{[:,l]}
&=\Big(2\beta \text{diag} \big((\mathbf{W}_{[:,l]})^{\circ (2)}\big)+\mathbf{A}^H\mathbf{A}\Big)^{-1}\mathbf{D}_{[:,l]} \nonumber \\
&=\Big(\mathbf{P}_l-\mathbf{P}_l\mathbf{A}^H(\mathbf{I}_{L}+
\mathbf{A}\mathbf{P}_l\mathbf{A}^H)^{-1}\mathbf{A}\mathbf{P}_l \Big)\mathbf{D}_{[:,l]}
\end{align}
for $l = 1,...,GM$, where $\mathbf{P}_l:=({0.5}/{\beta})\text{diag} \Big((\mathbf{W}_{[:,l]})^{\circ (-2)}\Big)$.



The $\mathbf{Z}$-subproblem is to minimize (\ref{d2}) with respect to $\mathbf{Z}$, which is given by
\begin{align}
\label{d6}
\min\limits_{\mathbf{Z}}\
\alpha_1 \sum_{i=1}^K \left \| \mathbf{Z}_i \right \|_F - \Re\big{\{}tr(\boldsymbol{\lambda}_1^H\mathbf{Z})\big{\}}+\frac{\beta}{2}\left \| \mathbf{Z}-\mathbf{W}\circ \mathbf{X} \right \|^2_F
\end{align}

Clearly, the minimization problem (\ref{d6}) can be solved separately for each row chunk $\mathbf{Z}_i$.
With a little algebra, (\ref{d6}) is equivalent to
\begin{align}
\label{d7}
\sum_{i=1}^K \ \min\limits_{\mathbf{Z}_i}\left [ \alpha_1 \left \| \mathbf{Z}_i \right \|_F
+\frac{\beta}{2}\left \| \mathbf{Z}_i-\mathbf{W}_i\circ \mathbf{X}_i
-\frac{1}{\beta_1}(\boldsymbol{\lambda}_1)_i
 \right \|^2_F
 \right ]
\end{align}

For the $i$th subproblem in (\ref{d7}), we show in Appendix B that it has a closed-form solution. Then the optimal solution of (\ref{d7}) is
\begin{align}
\label{d8}
\hat{\mathbf{Z}}_i=max\Big{\{}\left \| \widetilde{\mathbf{Z}}_i \right \|_F-\frac{\alpha_1}{\beta},\ 0\Big{\}}\frac{\widetilde{\mathbf{Z}}_i}{\left \| \widetilde{\mathbf{Z}}_i \right \|_F}, \ \forall  i = 1,\ ... \ ,K,
\end{align}
where $\widetilde{\mathbf{Z}}=\mathbf{W}\circ \mathbf{X}+(1/\beta) \boldsymbol{\lambda}_1$. To simplify expression, we denote the above chunk-wise soft shrinkage operation as
\begin{align}
\label{d8}
\hat{\mathbf{Z}}=Shrink_{(N,GM)}\big(\mathbf{W}\circ \mathbf{X}+\frac{1}{\beta} \boldsymbol{\lambda}_1,\frac{\alpha_1}{\beta} \big)
\end{align}
where $(N,GM)$ is the dimension of shrinkage operation in the objective matrix $\mathbf{W}\circ \mathbf{X}+\frac{1}{\beta} \boldsymbol{\lambda}_1$.


Similarly to $\mathbf{Z}$-subproblem, the $\mathbf{Q}$-subproblem also has a closed-form solution:
\begin{align}
\label{d9}
\hat{\mathbf{Q}}=Shrink_{(N,M)}\big(\mathbf{W}\circ \mathbf{X}+\frac{1}{\beta} \boldsymbol{\lambda}_2,\frac{\alpha_2}{\beta} \big)
\end{align}

At last, the multipliers matrix $\boldsymbol{\lambda}_1$ and $\boldsymbol{\lambda}_2$ are updated in standard way as following,
\begin{equation}
\label{d10}
\begin{cases}
\ \boldsymbol{\lambda}_1 \ {\leftarrow} \boldsymbol{\lambda}_1-\beta(\mathbf{Z}-\mathbf{W}\circ \mathbf{X})
\\
\ \boldsymbol{\lambda}_2 \ {\leftarrow} \boldsymbol{\lambda}_2-\beta(\mathbf{Q}-\mathbf{W}\circ \mathbf{X})
\end{cases}
\end{equation}

Note that there are three variables $\mathbf{X}$, $\mathbf{Z}$ and $\mathbf{Q}$ updated in our method, which is different from standard ADMM theory in which only two blocks of variables are updated alternatively. However, the functional (\ref{d2}) is separable in $\mathbf{Z}$ and $\mathbf{Q}$, i.e.,
$\min\limits_{\mathbf{Z},\mathbf{Q}}\mathcal{L}(\cdot )
=\min\limits_{\mathbf{Z}}\big\{\min\limits_{\mathbf{Q}}\mathcal{L}(\cdot )\big\}=\min\limits_{\mathbf{Q}}\big\{\min\limits_{\mathbf{Z}}\mathcal{L}(\cdot )\big\}$,
then the $\mathbf{Z}$-subproblem and $\mathbf{Q}$-subproblem in our method can be merged into one subproblem. Hence our method owns the same convergency as the standard ADMM \cite{bertsekas1989parallel}.
In short, we have the following result.
\begin{theorem}
\label{theorem2}
For any $\beta,\alpha_1,\alpha_2>0$, and fixed $\mathbf{W}$,
the iteration of $\mathbf{X}$, $\mathbf{Z}$ and $\mathbf{Q}$ in the ADMM from
any initial point is guaranteed to converge to a global minimizer of (\ref{d1}). Specially, $\mathbf{X}$ converges to a solution of (\ref{c3}).
\end{theorem}

We iterate the ADMM for several times, in each of which we update $\mathbf{W}$ by (\ref{d11}) using the previous estimation of $\mathbf{X}$ as the same as in SOCP.
For convenience, we summarize our ADMM in Table \ref{admmtable}.

\begin{table}[!htp]
\setlength{\abovecaptionskip}{-9pt}
\setlength{\belowcaptionskip}{-6pt}
\caption{ADMM}
\label{admmtable}
\begin{algorithm}[H]
\caption*{{\bf Algorithm} Alternating Direction Method of Multipliers}
{\normalsize
\begin{algorithmic}[1]
\STATE {Initialize variables $\mathbf{Z}$, $\mathbf{Q}$,
weight matrix $\mathbf{W}$, and multipliers $\boldsymbol{\lambda}_1$, $\boldsymbol{\lambda}_2$.}
\STATE {Initialize parameters $\alpha_1$, $\alpha_2$, $\beta$, $\epsilon$ and the iteration number $MaxCount$. Set the iteration counter $count \leftarrow 1$.}
 \WHILE{$count \leq MaxCount$ }
 \WHILE{not convergent and stopping criterion not met}
  \STATE {Update $\mathbf{X}$ with (\ref{d5}), $\mathbf{Z}$ with (\ref{d8}), $\mathbf{Q}$ with (\ref{d9}), in turn.}
  \STATE {Update $\boldsymbol{\lambda}_1$ and $\boldsymbol{\lambda}_2$ with (\ref{d10}).}
 \ENDWHILE
 \STATE {Update $\mathbf{W}$ with (\ref{d11}).}
 \STATE {$count \leftarrow count + 1.$}
 \ENDWHILE
\end{algorithmic}}
\end{algorithm}
\end{table}

%

\section{Numerical Results}
In this section, we carry out experiments to illustrate
the performances of our proposed functional and efficient algorithm.
Our SOCP and ADMM for solving functional (\ref{c3}) are addressed as ``SolveSOCP'' and ``SolveADMM'' respectively in this simulation.
As only the sparsity of user activities is exploited if we eliminate the second term in (\ref{c3}) by setting $\alpha_2=0$, we address it ``SolveRowLasso'' and add it into comparison.
Correspondingly, we eliminate the first term in (\ref{c3}) by setting $\alpha_1=0$, and address it ``SolveElementLasso'' which only exploits the sparse relationship between active users and effective RRHs surrounding it.
SolveRowLasso and SolveElementLasso can be considered just as special cases of our proposed functional, and both of them are solved by SOCP in this simulation.

The method ``modified Bayesian compressive sensing'' (``BCS'' for short) in \cite{xu2015active} is also included for comparison.
BCS is based on the exact knowledge of large-scale path loss between each RRH and each user, which, however, is costly in real C-RAN and may not be predicted accurately due to mobility and environment changes.
Here we include another case named ``BCS2'' in which BCS is provided with inaccurate large-scale path loss parameters.
We address BCS and BCS2 as ``SolveBCS'' and ``SolveBCS2'' respectively in this simulation.

We set the number of RRHs $G=10$ and number of users $K=100$.
The number of antennas in each RRH and user are $M=3$ and $N=2$ respectively.
Rayleigh fading channel model is assumed between RRHs and users.
The size of active user set $\mathcal{A}$ is $10$.
SNR is set to $10$ dB.
Each element of pilot matrix $\mathbf{P}$ in (\ref{b4}) has an
independent and identical normalized complex gaussian distribution.

We empirically set $\alpha_1$ and $\alpha_2$ in functional (\ref{c3}) equal to a small percentage of $\alpha_1^*$ and $\alpha_2^*$ respectively, say 1\%-5\%.
$\beta$ in the ADMM functional (\ref{d1}) has the same value level as $\alpha_1$ and $\alpha_2$ to accelerate convergence.
The number of re-weighting times $MaxCount$ in both SOCP and ADMM is set to 2.
$\epsilon$ in (\ref{d11}) is set to $10^{-8}$.
Assuming the precise large-scale path loss between $k$th user and $g$th RRH is
$\tau_{g,k}$, we provide BCS2 with $\tau_{g,k}\cdot(1 + 0.5n)$ where $n$ obeys  standard normal distribution. All the results are averaged over 50 runs.

At first, we demonstrate the channel estimation results of respective algorithms, as shown in Fig. 1
where NMSE denotes the normalized mean squared error.
It can be seen that SOCP and BCS provide the best performances.
Because of the tradeoff between preciseness and efficiency in the implementation, our ADMM works slightly worse than BCS and SOCP, but still lies in the same level.
The performances of ``ElementLasso'' and ``RowLasso'' are worse than SOCP which is in accordance with our expectation.
BCS2 has a poor performance for reliance on the exact information of large-scale path loss parameters.
\vspace*{-3pt}
\begin{figure}[h!]
\setlength{\abovecaptionskip}{-1pt}
\setlength{\belowcaptionskip}{3pt}
\centering
\includegraphics[angle=0,scale=0.4]{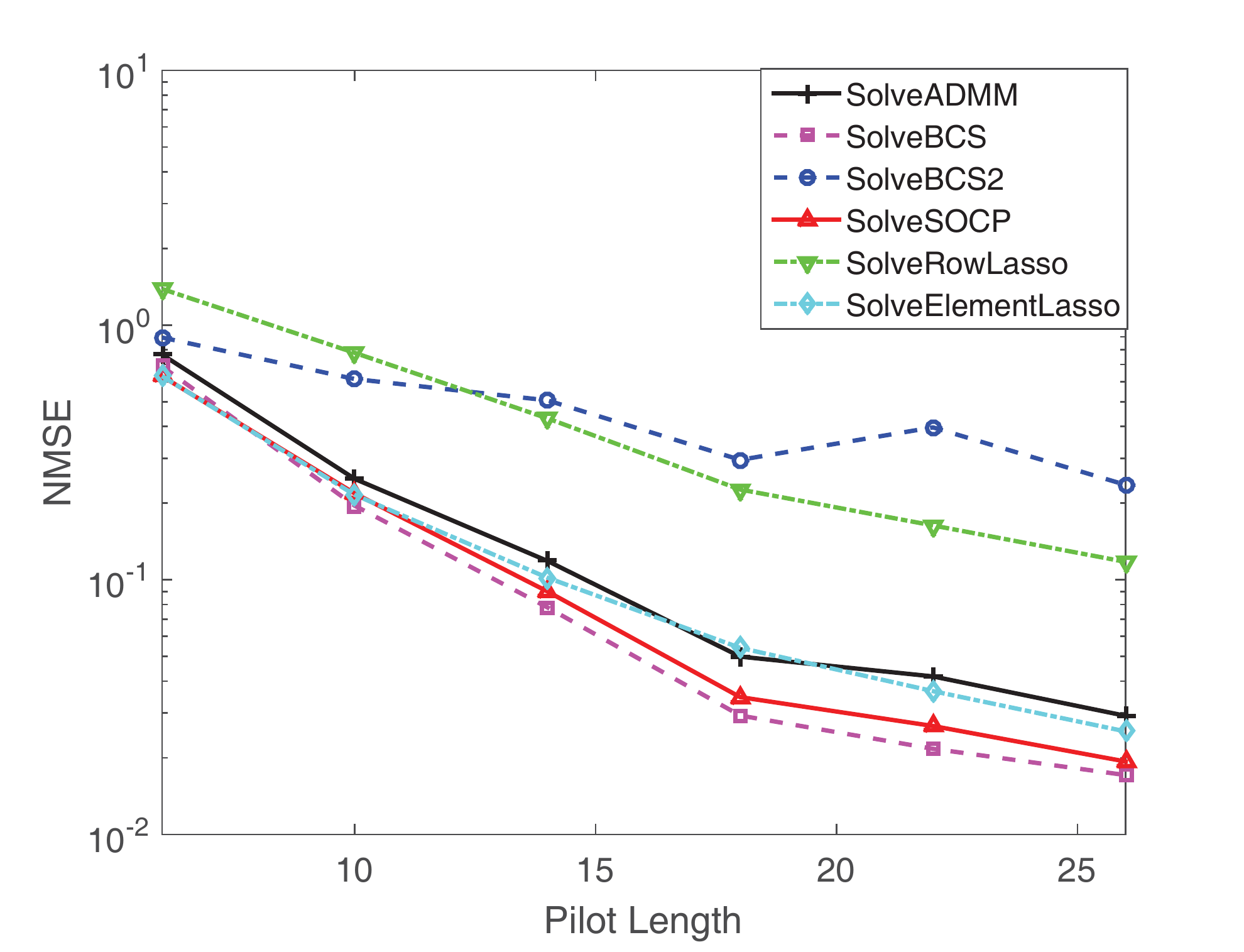}
\caption{NMSE of respective algorithms vs. pilot length}
\end{figure}
\setlength{\textfloatsep} {3pt}

\begin{figure}[h!]
\setlength{\abovecaptionskip}{-1pt}
\setlength{\belowcaptionskip}{3pt}
\centering
\includegraphics[angle=0,scale=0.4]{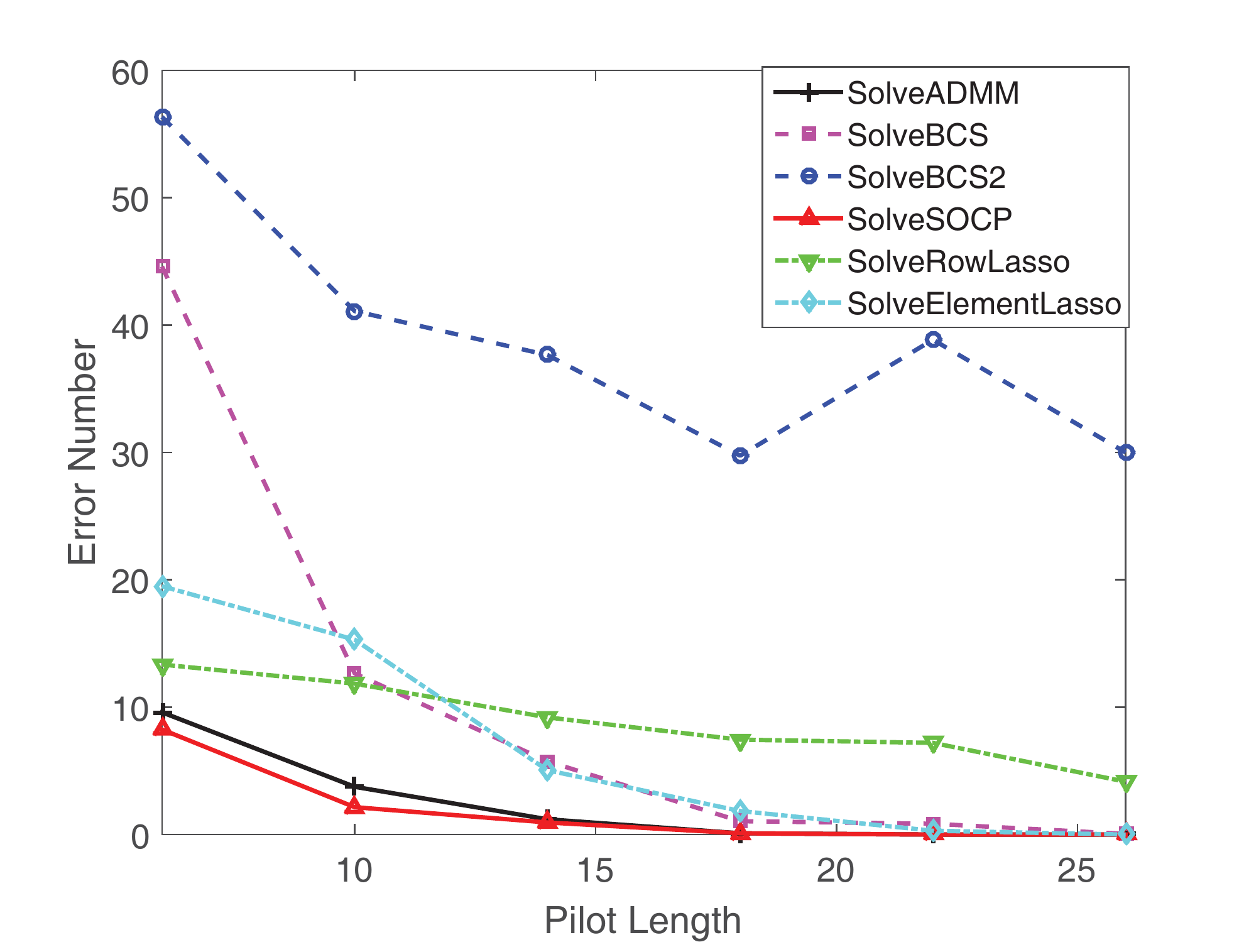}
\caption{Multi-user detection results of respective algorithms vs. pilot length}
\end{figure}
\setlength{\textfloatsep} {3pt}

Fig. 2 shows the wrongly detected user number of respective algorithms vs. pilot length.
We can see that our SOCP and ADMM exhibit the best performances in detecting activities of users. ElementLasso, RowLasso and BCS need longer training pilots to detect users precisely.
It can be observed from Fig. 3 that ADMM and BCS have the lowest computational complexity.
SOCP has a rather high complexity particularly when pilot is long,
which attributes to the increase of problem dimension.
The complexity of ADMM and BCS do not increase with pilot length
within our simulation range
because that properly longer pilot helps them to converge faster.

\begin{figure}[h!]
\setlength{\abovecaptionskip}{-1pt}
\setlength{\belowcaptionskip}{3pt}
\centering
\includegraphics[angle=0,scale=0.4]{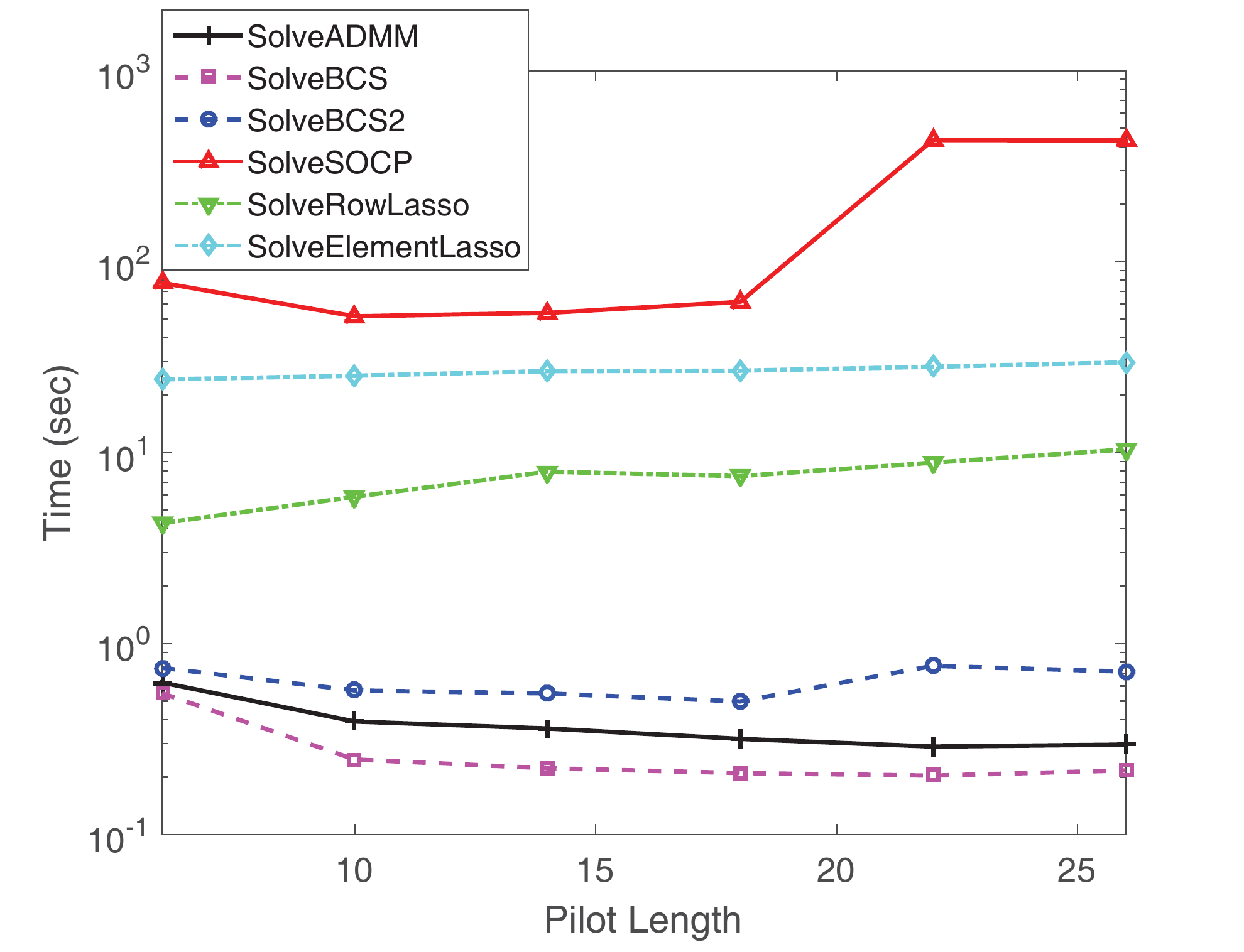}
\caption{Computation time of respective algorithms vs. pilot length}
\end{figure}
\setlength{\textfloatsep} {3pt}

%
%
%

%


\section{Conclusion}
We proposed a mixed $\ell_{2,1}$-regularization functional to solve  channel estimation and multi-user detection problems in C-RAN.
This functional simultaneously exploits the sparsity of user activities and sparsity of RRHs related to each active user.
Guidelines are also provided to help choosing tuning parameters.
To accelerate the processing procedure, an efficient method based on ADMM is proposed. Numerical results show that our functional solved by our efficient method achieves state-of-the-art performance and also has low computational complexity.

%
%

\section{Appendix}
\subsection{Proof of Theorem \ref{Theorem1}}

At first we prove the solution of (\ref{c3}) $\hat{\mathbf{X}}$ equals to zero if $\alpha_1\geq \alpha_1^*=\max\limits_{i} \left \| (\mathbf{A}^H_i\mathbf{B}) \circ (\mathbf{W}_i)^{\circ (-1)} \right \|_F$.

To simplify expressions, we denote the three positive terms in (\ref{c3}) as $a_1(\mathbf{X})$, $a_2(\mathbf{X})$ and $c(\mathbf{X})$ in turn.
It can be seen that
\begin{align}
\label{f1}
\min\limits_{\mathbf{X}}\big(a_1(\mathbf{X})+c(\mathbf{X})\big)+
\min\limits_{\mathbf{X}'}a_2(\mathbf{X}')
\leq (\ref{c3})
\end{align}
where equality holds when the minimum solutions $\hat{\mathbf{X}}=\hat{\mathbf{X}'}$.

As $a_1(\mathbf{X})+c(\mathbf{X})$ is convex, by taking derivative, we get
\begin{equation}
\label{c4}
\mathbf{A}^H (\mathbf{A}\mathbf{X}-\mathbf{B})+\alpha_1 \mathbf{V}=\mathbf{0}
\end{equation}
where $\mathbf{V}$ is the derivative of $\sum_{i=1}^K \left \| \mathbf{W}_i \circ \mathbf{X}_i \right \|_F$ with regard to $\mathbf{X}$, and each $\mathbf{V}_i$ satisfies
\begin{equation}
\label{c5}
\mathbf{V}_i=\begin{cases}
\frac{\mathbf{X}_i \circ (\mathbf{W}_i)^{\circ (2)}}{\left \| \mathbf{X}_i  \circ \mathbf{W}_i \right \|_F} & \text{ iff } \mathbf{X}_i\neq \mathbf{0} \\
\in \{ \mathbf{V}_i:\left \| \mathbf{V}_i \circ (\mathbf{W}_i)^{\circ (-1)}\right \|_F\leq 1  \} & \text{ iff } \mathbf{X}_i= \mathbf{0}
\end{cases}
\end{equation}

Hence the solution of $\min\limits_{\mathbf{X}}\big(a_1(\mathbf{X})+c(\mathbf{X})\big)$ is zero iff
$ \left \| (\mathbf{A}^H_i\mathbf{B}) \circ  (\mathbf{W}_i)^{\circ (-1)} \right \|_F \leq {\alpha_1},\ \forall k=1,...,K $.
The solution of $\min\limits_{\mathbf{X}'}a_2(\mathbf{X}')$ is zero.
Therefore zero is also the solution of (\ref{c3}) when $\alpha_1\geq \max\limits_{i} \left \| (\mathbf{A}^H_i\mathbf{B})  \circ (\mathbf{W}_i)^{\circ (-1)} \right \|_F$.

The proof that solution of (\ref{c3}) equals to zero if $\alpha_2\geq \alpha_2^*=\max\limits_{i,j} \left \| (\mathbf{A}^H_i\mathbf{B}_{j})  \circ (\mathbf{W}_{i,j})^{\circ (-1)} \right \|_F$ can be derived similarly.

\subsection{Complex Matrix Shrinkage}
The shrinkage operator for complex matrix is a direct extension of the basic one-dimensional soft thresholding method or shrinkage in, e.g., \cite{parikh2014proximal}. To find the solution of an optimization problem as blow
\begin{align}
\label{f0}
\min\limits_{\mathbf{X}}\ \alpha \left \| \mathbf{X} \right \|_F
+\frac{\beta}{2}\left \| \mathbf{X}-\mathbf{B} \right \|^2_F
\end{align}
where $\mathbf{X}, \ \mathbf{B} \in \mathbb{C}^{N\times M}$. As (\ref{f0}) is convex but non-differentiable, by taking the derivative of it, we get
\begin{align}
\label{f0-1}
\alpha \mathbf{V} + \beta({\mathbf{X}}-\mathbf{B}) = 0
\end{align}
where $\mathbf{V}\in \mathbb{C}^{N\times M}$ is the subgradient of
$\left \| {\mathbf{X}} \right \|_F$
, and
\begin{equation}
\label{f0-2}
\mathbf{V}=\begin{cases}
\frac{{\mathbf{X}}}{\left \| {\mathbf{X}} \right \|_F} & \text{ iff } {\mathbf{X}}\neq \mathbf{0} \\
\in \{ \mathbf{V}:\left \| \mathbf{V} \right \|_F\leq 1  \} & \text{ iff } {\mathbf{X}}= \mathbf{0}
\end{cases}
\end{equation}

By plugging (\ref{f0-2}) into (\ref{f0-1}), we get the optimal solution
$\hat{\mathbf{X}}=(\left \|\mathbf{B}\right \|_F-{\alpha}/{\beta}){\mathbf{B}}/{\left \|\mathbf{B}\right \|_F}$
iff $\hat{\mathbf{X}}\neq \mathbf{0}$, and $\left \|\mathbf{B}\right \|_F\leq {\alpha}/{\beta}$ iff $\hat{\mathbf{X}}=\mathbf{0}$.
At last, we arrive at
\begin{align}
\label{f0-3}
\hat{\mathbf{X}}=max\{\left \|\mathbf{B}\right \|_F-\frac{\alpha}{\beta},0\}\frac{\mathbf{B}}{\left \|\mathbf{B}\right \|_F}
\end{align}

%

\bibliographystyle{IEEEtran}

\end{document}